\documentclass[prb,twocolumn,superscriptaddress,showpacs,floatfix]{revtex4}

\usepackage{epsfig,amsmath,amssymb,latexsym}

\usepackage{graphicx}
\usepackage{subfigure}

\begin{document}

\title{Electronic structure of relativistic Mott insulator Li$_2$RhO$_3$} \preprint{1}

\author{Chao Cao}
 \email[E-mail address: ]{ccao@hznu.edu.cn}
 \affiliation{Condensed Matter Group,
  Department of Physics, Hangzhou Normal University, Hangzhou 310036, China}
\author{Yongkang Luo}
\author{Zhuan Xu}
  \affiliation{Department of Physics,
  Zhejiang University, Hangzhou 310027, China}
\author{Jianhui Dai}
  \affiliation{Condensed Matter Group,
  Department of Physics, Hangzhou Normal University, Hangzhou 310036, China}

\date{Feb. 1, 2013}

\begin{abstract}
Motivated by studies of coexisting electron correlation and spin-orbit coupling effect in Na$_2$IrO$_3$ and a recent experiment of its 4d analogue Li$_2$RhO$_3$, we performed first-principles calculations of the rhodium oxide compound. The experimentally observed ground state of Li$_2$RhO$_3$ can be recovered only if both spin-orbit coupling and on-site Coulomb interaction are taken into consideration. Within the proper $U$ range for 4d-orbitals ($2\leqslant U\leqslant 4$ eV), the ground state of Li$_2$RhO$_3$ could be either zigzag-AFM or stripy-AFM, both yielding energy gap close to experimental observation. Furthermore, the total energy differences between the competing magnetic phases are $\leqslant 3$ meV/Rh within $2\leqslant U\leqslant 4$ eV, manifesting strong magnetic frustration in the compound. Finally, the phase energy of Li$_2$RhO$_3$ cannot be fitted with the two-dimensional Heisenberg-Kitaev model involving only the nearest neighbor interactions, and we propose that inter-layer interactions may be responsible for the discrepancy.
\end{abstract}

\maketitle

  Electronic structure of transition metal oxides has been a hot topic in the condensed matter community for decades. On one hand, the 3d-transition metal oxides are considered as typical strongly correlated systems whose on-site electron-electron interactions $U$ are large enough to forbid the hopping of electrons among lattice sites, leading to the Mott behavior. On the other hand, as the 5d-orbitals are much more extended than the 3d-orbitals, the correlation effect for 5d-orbitals are small and negligible, while its spin-orbit coupling (SOC) effect becomes prominent. 
  
  The above argument, however, is challenged by studies of 5d-metal oxide Na$_2$IrO$_3$\cite{PhysRevLett.110.076402,PhysRevB.85.180403,PhysRevLett.109.197201,PhysRevLett.109.266406,PhysRevB.83.245104,PhysRevLett.102.256403,PhysRevLett.108.127203,PhysRevLett.108.127204,PhysRevLett.108.106401,PhysRevLett.110.097204}. As a 5d metal oxide, Na$_2$IrO$_3$ is an ideal candidate for realizing the Heisenberg-Kitaev model 
  \begin{equation}
    H_{\mathrm{HK}}=\sum_{<i,j>}\left[(1-\alpha)\hat{\mathbf{S}}_i\cdot \hat{\mathbf{S}}_j-2\alpha S^{\gamma}_i S^{\gamma}_j\right]
    \label{eq:hk}
  \end{equation}
  where $\gamma=x,y,z$ denotes the three different types of links in the hexagonal lattices, and $\alpha$ indicates the interpolation between the Heisenberg term and the Kitaev term, with $\alpha=0$ and $\alpha=1$ recovering the original Heisenberg and Kitaev model, respectively. It is believed that $\alpha$ is relatively large for 5d electron systems due to strong SOC effect. As pointed out by J. Reuther {\it et al.}, the ground state of the Heisenberg-Kitaev model is N\'{e}el antiferromagnetic (AFM) for $0\leqslant \alpha < 0.4$, and stripy AFM for $0.4\leqslant \alpha < 0.8$. When the Kitaev term becomes dominantly large where $0.8\leqslant\alpha<1$, the system will enter a spin-liquid phase\cite{PhysRevB.84.100406}. However, it was later discovered in both first-principles calculations and experiments that the ground state for Na$_2$IrO$_3$ is indeed the zigzag AFM phase, and the material is a relativistic Mott insulator\cite{PhysRevLett.109.266406,PhysRevB.85.180403}. To resolve the apparent discrepancy, it was proposed that a next-nearest-neighbor Heisenberg coupling term should be added into the original Heisenberg-Kitaev model\cite{PhysRevLett.108.127203}. On the contrary, I. I. Mazin {\it et al.} questioned the application of localized spin models in the system, and proposed a quasi-molecular-orbital approach\cite{PhysRevLett.109.197201}.
 
 \begin{figure}[htp]
  \centering
  \subfigure[Geometry] {
    \scalebox{0.14}{\includegraphics{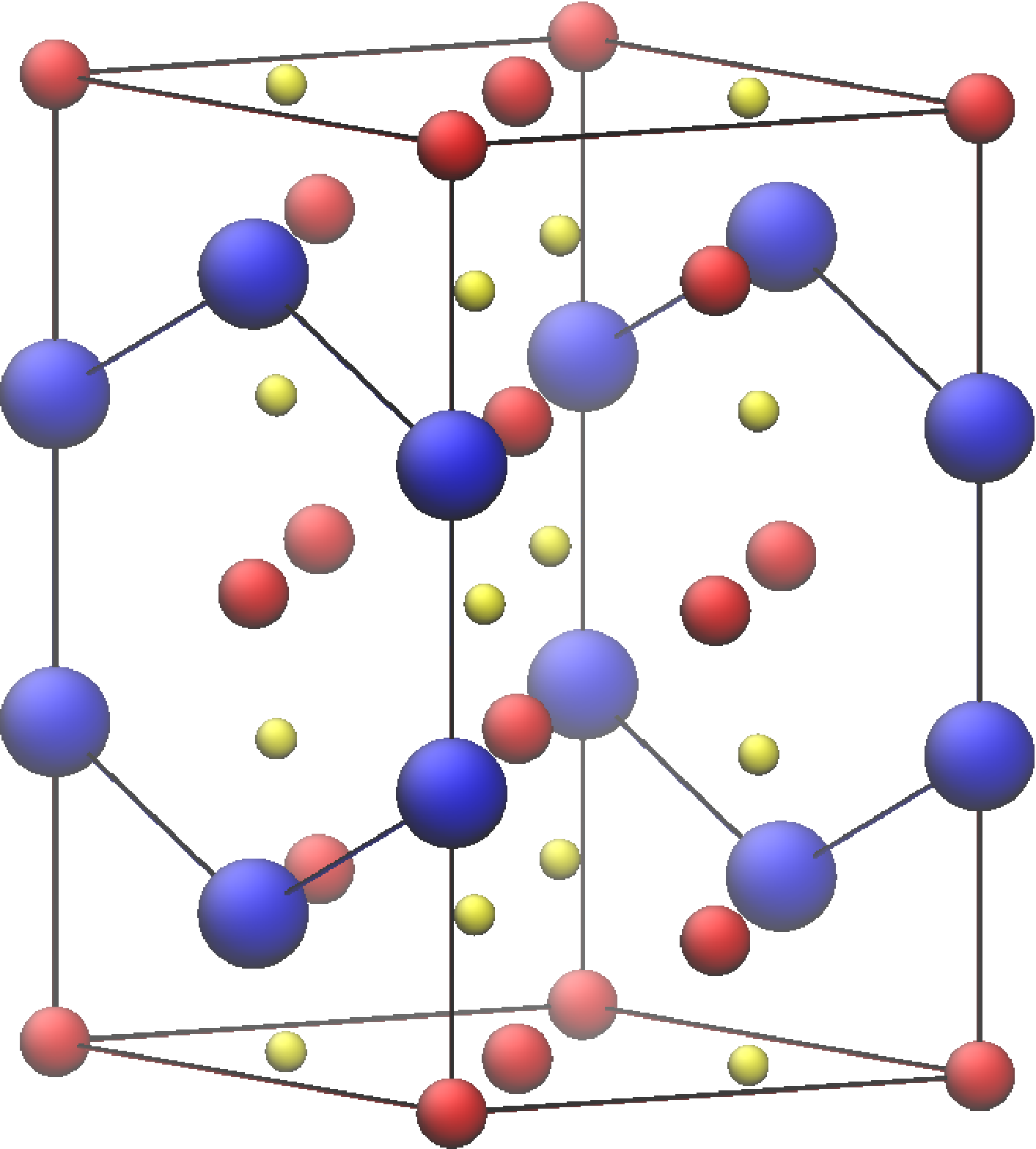}}
    \label{fig:geometry}}
  \subfigure[K-points] {
    \scalebox{0.25}{\includegraphics{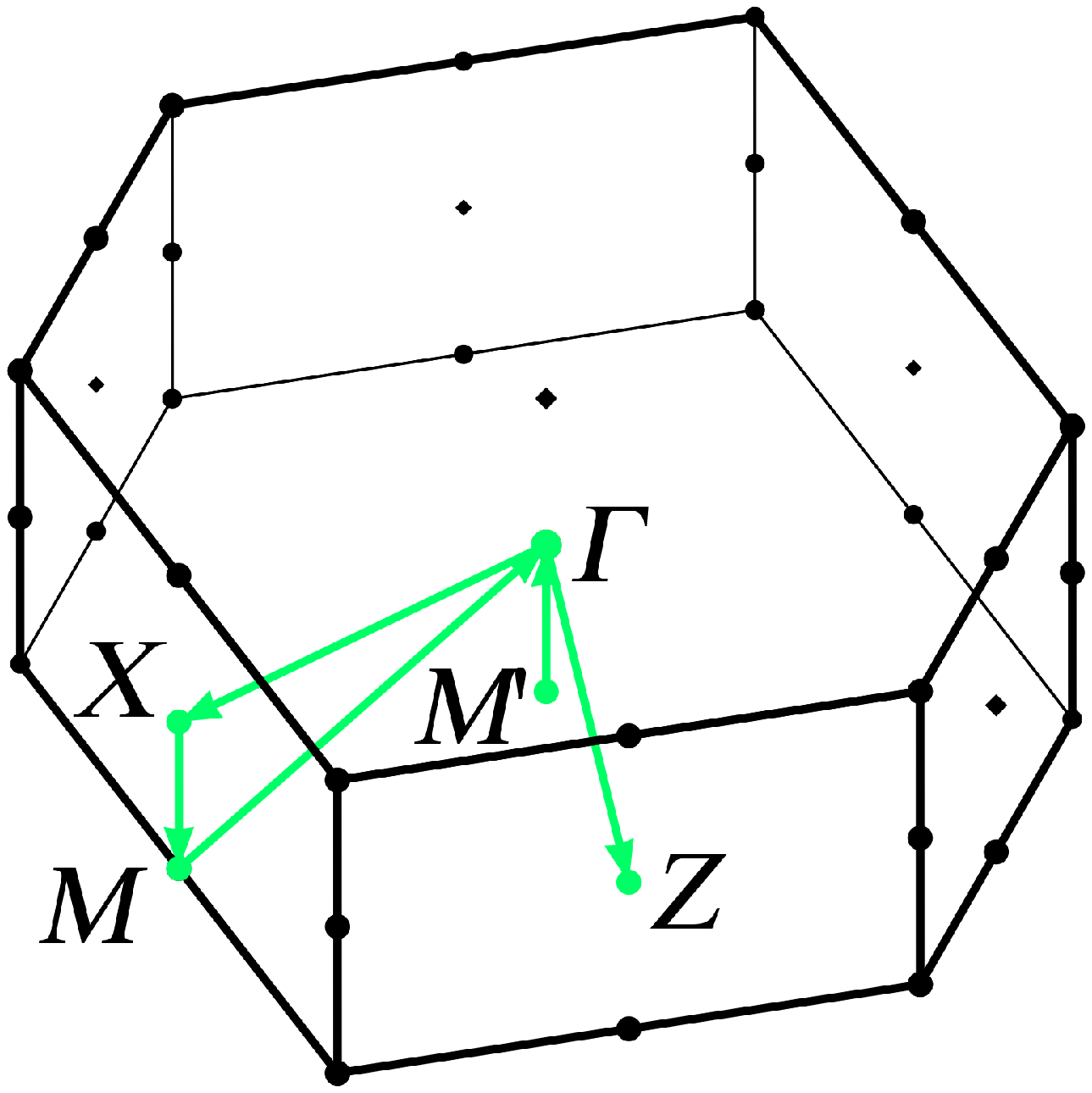}}
    \label{fig:kpoints}}
  \subfigure[N\'{e}el-AFM] {
    \scalebox{0.12}{\includegraphics{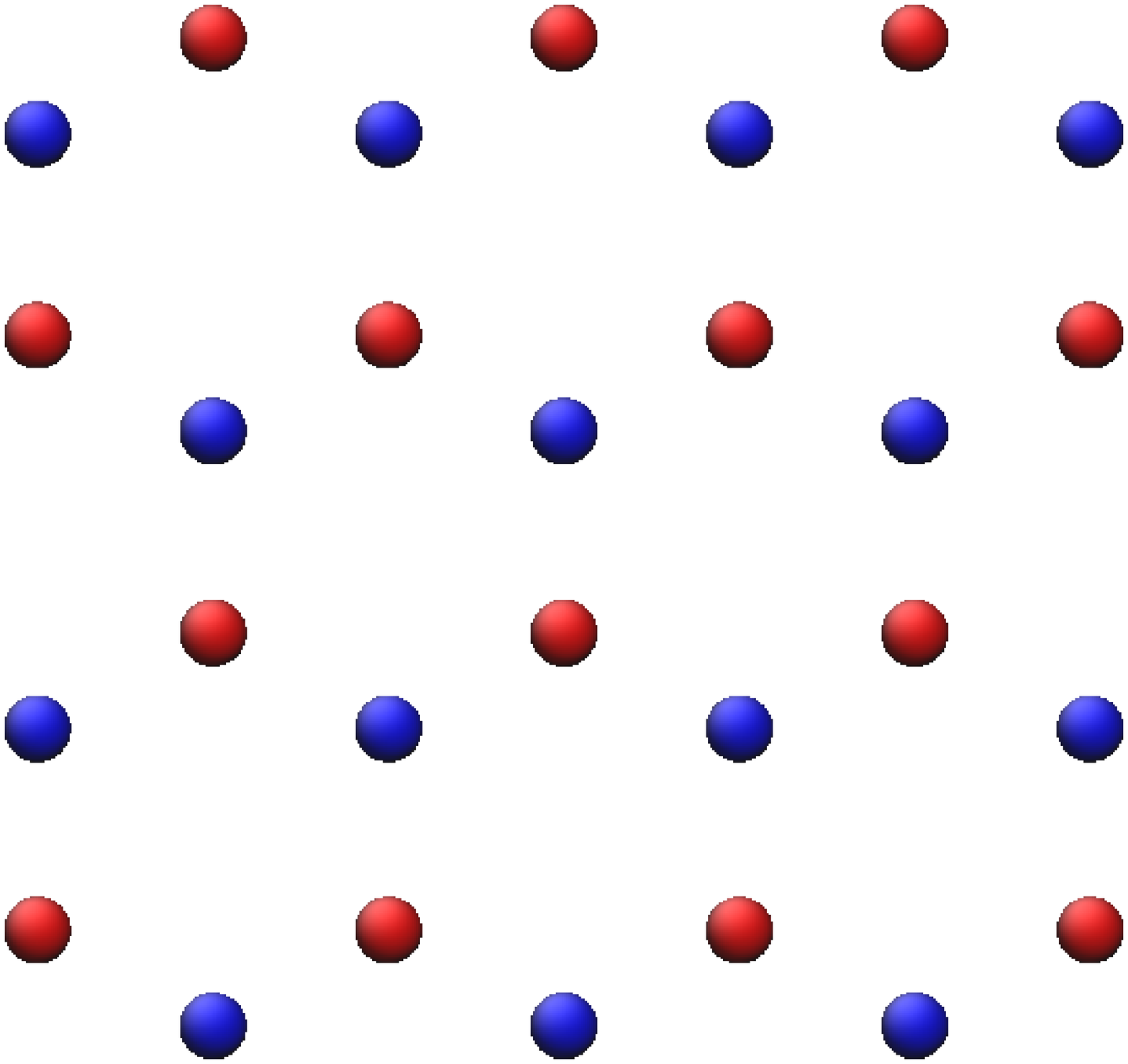}}
    \label{fig:neel}}
  \subfigure[Zigzag-AFM] {
    \scalebox{0.12}{\includegraphics{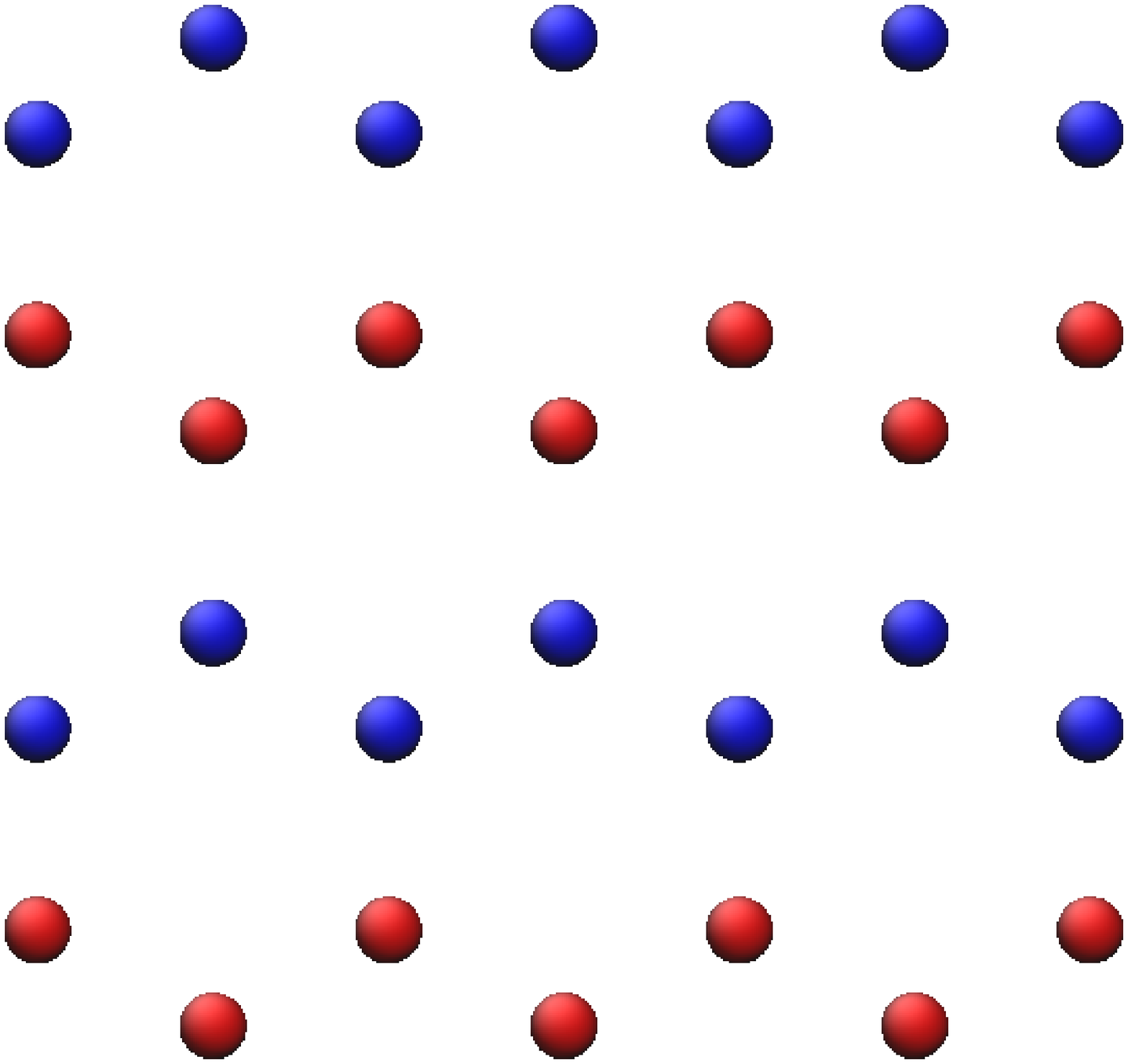}}
    \label{fig:zigzag}}
  \subfigure[Stripy-AFM] {
    \scalebox{0.12}{\includegraphics{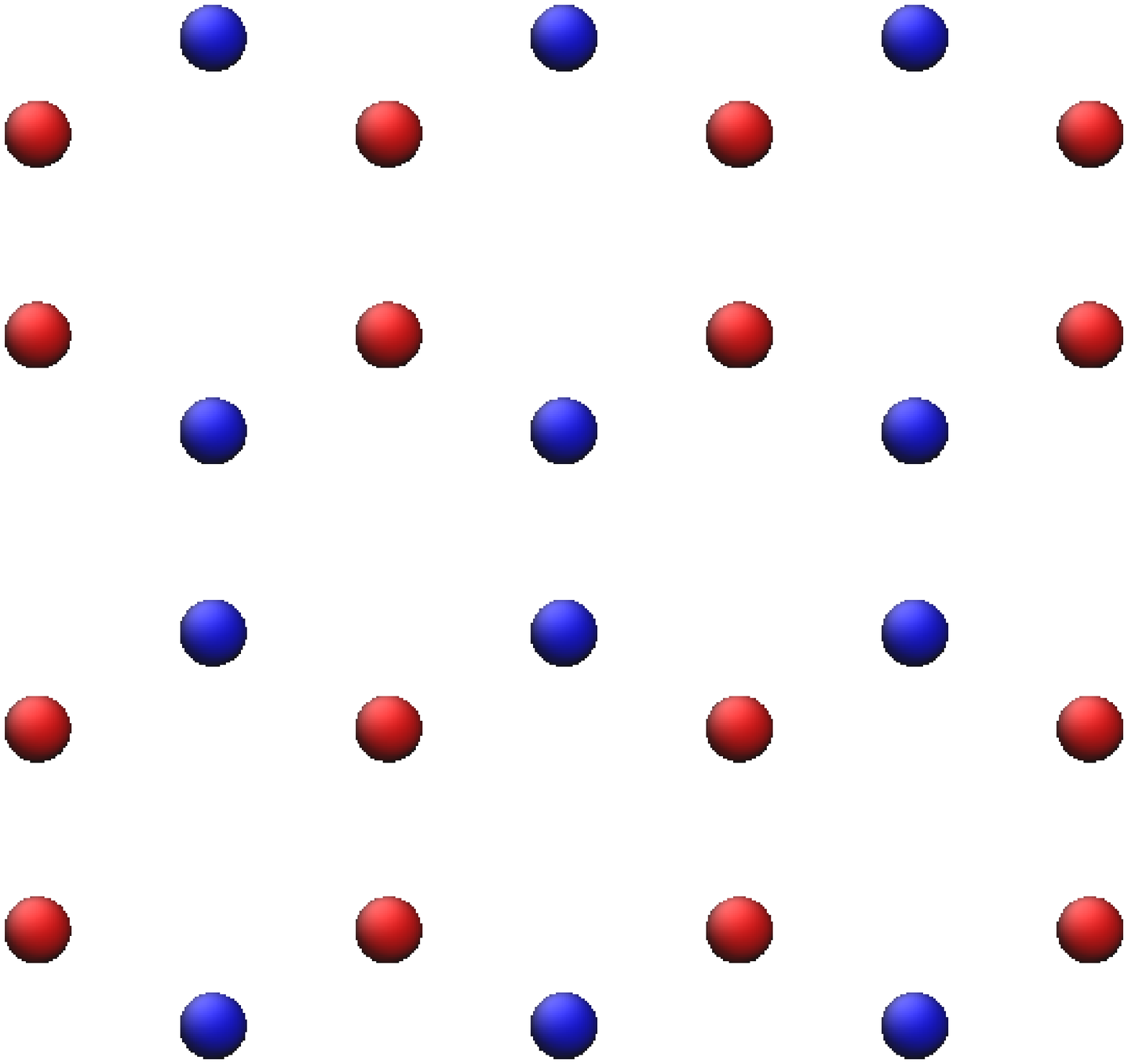}}
    \label{fig:stripy}}
  \caption{(a) Crystal structure of Li$_2$RhO$_3$. The blue large atoms are Rh; red smaller atoms are Li; and yellow smallest atoms are O. The hexagonal lattice of Rh is also shown. (b) The first irreducible Brillioun zone and the high symmetry points appeared in this paper. (c-e) The N\'{e}el-AFM, zigzag-AFM, and stripy-AFM configurations for hexagonal lattices. The red and blue spheres indicates spin up and down Rh atoms, respectively.}
 \end{figure}
  
\begin{table*}[ht]
  \begin{tabular}{c||c|c|c|c|c||c|c|c|c|c}
    \hline
     $U$(eV) & NM & N\'{e}el & Stripy & Zigzag & FM & NM$^{\mathrm{soc}}$ & N\'{e}el$^{\mathrm{soc}}$ & Stripy$^{\mathrm{soc}}$ & Zigzag$^{\mathrm{soc}}$ & FM$^{\mathrm{soc}}$ \\
     \hline
     0.0 & -138.1672 & -138.1672 & -138.2221 & -138.1994 & {\it -138.3524} & -138.8865 & -138.8867 & -138.9260 & -138.9170 & {\bf -139.0421} \\
     1.0 & -134.5490 & -134.6043 & -134.6979 & -134.6441 & {\it -134.8125} & -135.2827 & -135.4302 & -135.4407 & -135.4610 & {\bf -135.5143} \\
     2.0 & -131.0182 & -131.3321 & -131.3295 & -131.2195 & {\it -131.3751} & -131.7448 & -132.2128 & -132.2182 & {\bf -132.2294} & -132.2102 \\
     3.0 & -127.5755 & -128.2610 & -128.2703 & -128.1069 & {\it -128.2904} & -128.2831 & -129.1358 & -129.1415 & {\bf -129.1459} & -129.1332 \\
     4.0 & -124.1958 & -125.2890 & -125.3068 & -125.1276 & {\it -125.3149} & -124.8964 & -126.1658 & {\bf -126.1723} & -126.1492 & -126.1652 \\
     \hline
  \end{tabular}
  \caption{Total energies (per unit cell, four Ru atoms) of different Li$_2$RhO$_3$ magnetic phases. Columns 2-6 are for the calculations without SOC; while columns 7-11 are for the calculations with SOC. The lowest phase energy at different $U$ with and without SOC are indicated with italian and bold fonts, respectively.}
  \label{tab:mag_energy}
\end{table*}

 \begin{figure*}[t]
  \centering
  \subfigure[NM $U=0$ Bands] {
    \rotatebox{270}{\scalebox{0.5}{\includegraphics{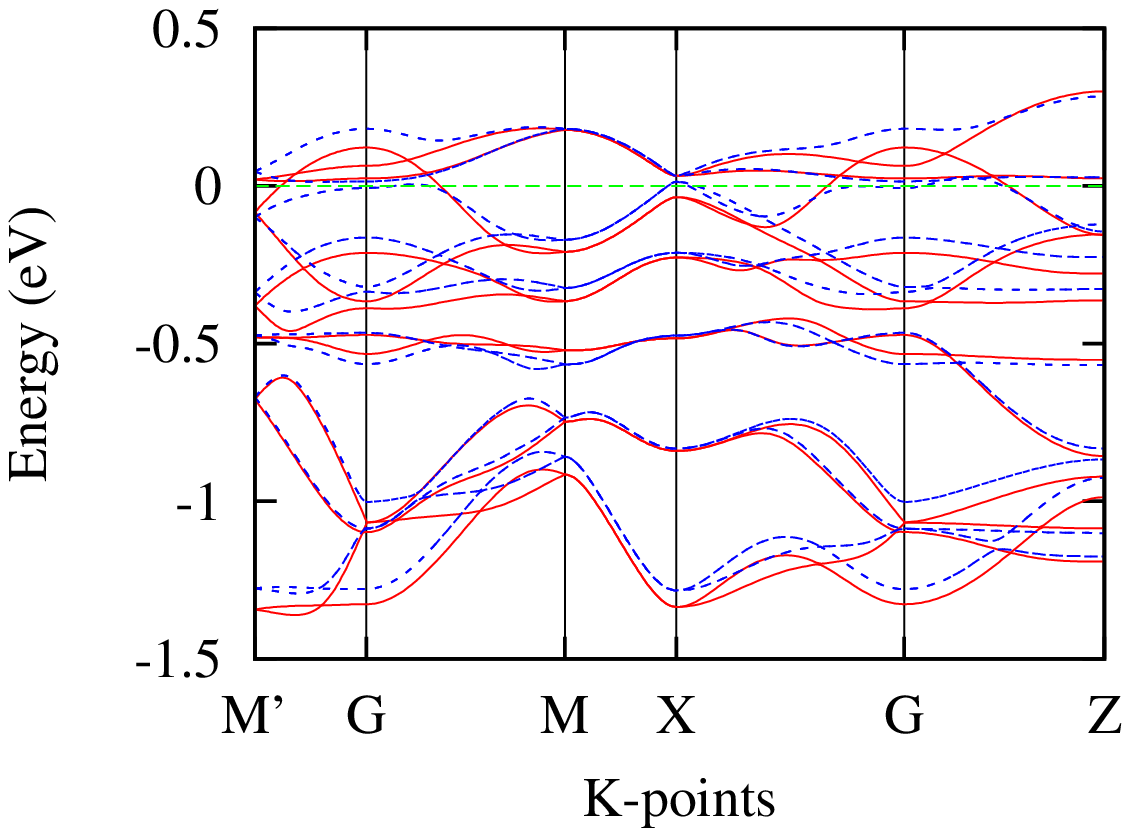}}}
    \label{fig:bs_nm}}
  \subfigure[NM $U=4$ Bands] {
    \rotatebox{270}{\scalebox{0.5}{\includegraphics{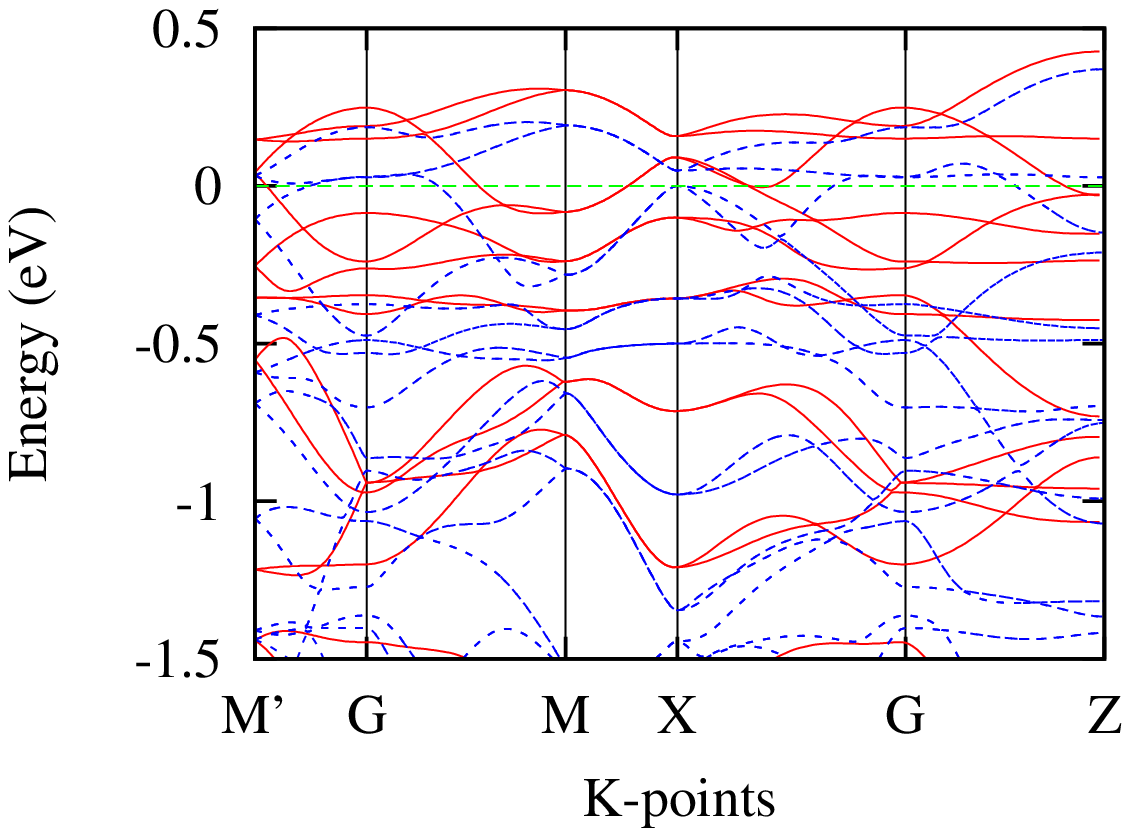}}}
    \label{fig:bs_nm_soc}}
  \subfigure[NM SOC+$U=0$ DOS] {
    \rotatebox{270}{\scalebox{0.5}{\includegraphics{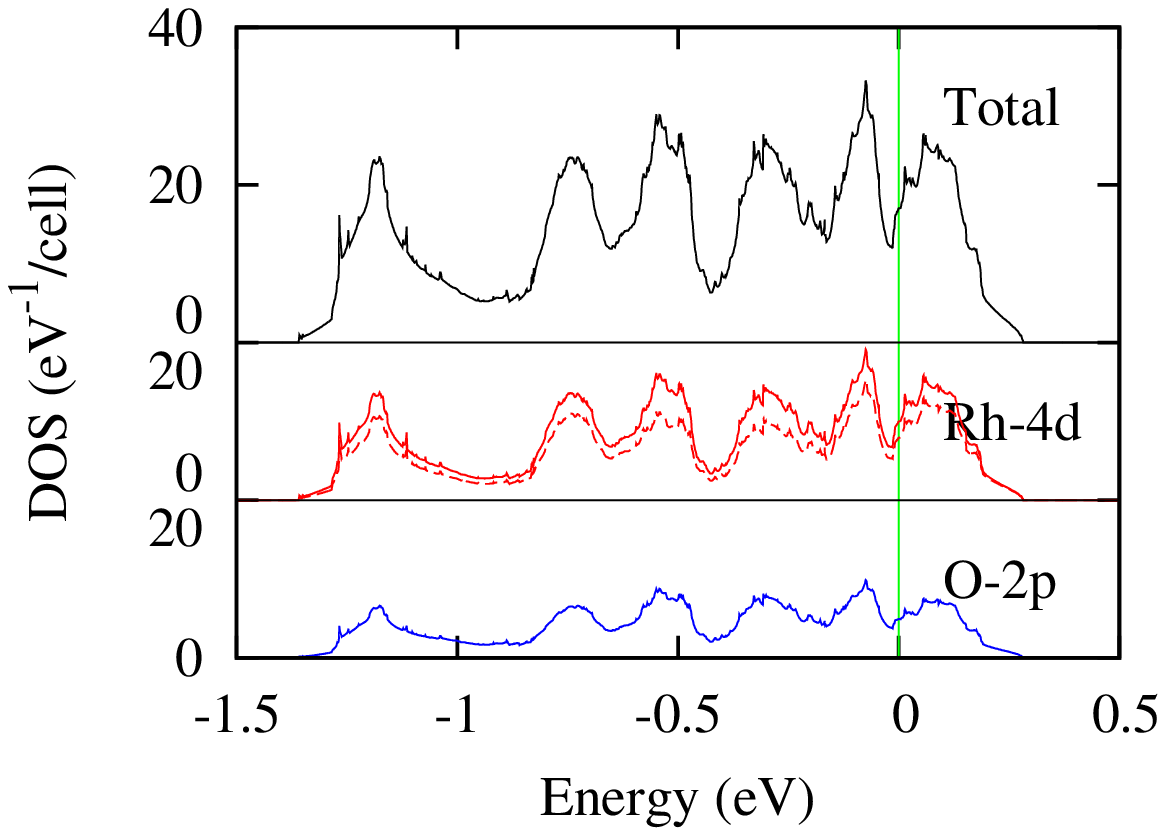}}}
    \label{fig:dos_nm}}
  \subfigure[NM SOC+$U=4$ DOS] {
    \rotatebox{270}{\scalebox{0.5}{\includegraphics{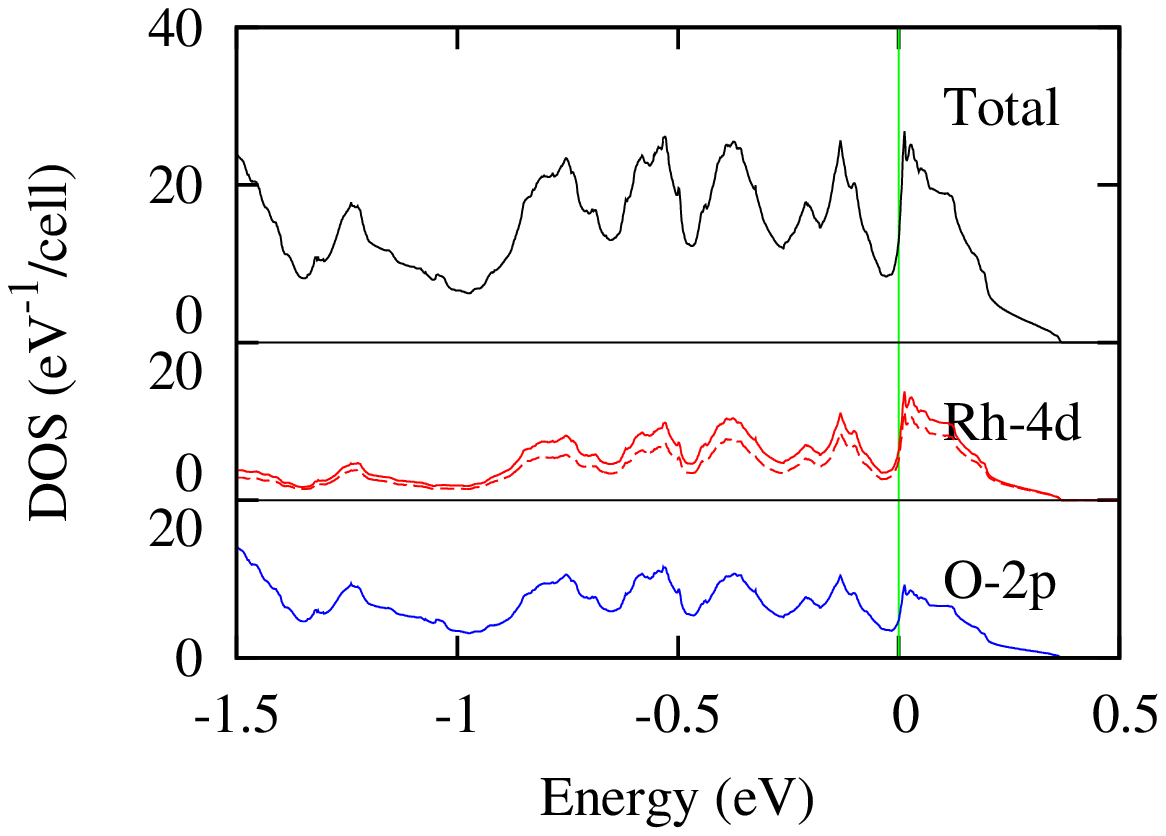}}}
    \label{fig:dos_nm_soc}}

  \caption{(a-b): Band structure of NM phase without (red solid lines) and with (blue dashed lines) SOC. (c-d): Total and partial DOS of NM phase with SOC. (a) and (c): $U$=0 eV. (b) and (d): $U$=4 eV.\label{fig:bsdos_nm_u0}}
 \end{figure*}

  Recently, 4d-transition metal oxide Li$_2$RhO$_3$ was reported to be in close proximity to spin glassy phase\cite{2013arXiv1303.1235L}. As the 4d systems usually have moderate Coulomb interaction and SOC in between 3d and 5d systems, this 4d compound should be very interesting not only because of its structural analogue to Na2IrO3, but also due to the more delicate interplay between the Coulomb correlation and SOC. It was determined experimentally that no long-range magnetic order was observed in Li$_2$RhO$_3$ down to 0.5 K, and spin-freezing temperature was measured to be around 6 K. Fitting of the resistivity data suggested that the compound was a semiconductor with a narrow band gap of $\sim$78 meV. It is worthy noting that anti-site disorder effect between Li$^+$ and Rh$^{4+}$ ions is inevitable in the experiment, therefore it is yet to be determined whether the spin-glassy feature is intrinsic or due to the anti-site defects. The experiment therefore raises several interesting questions: 1. Is the narrow band gap due to electron correlation or spin-orbit coupling effect? 2. How close is the defect-free Li$_2$RhO$_3$ to the spin-glassy phase? 3. How relevant is the system to the Heisenberg-Kitaev model?

  In this article, we present our latest first principles study of the 4d-transition metal oxide Li$_2$RhO$_3$. We calculated and compared the total energy of its magnetic ordered phases with and without SOC effect as well as for different on-site interaction $U$. We analyzed its band structure and density of states (DOS) at different $U$'s with and without SOC. We conclude that, both the SOC and on-site interaction $U$ are important in the compound, and the energy gap is indeed a correlation effect. Moreover, the magnetism in Li$_2$RhO$_3$ is extremely frustrated, and thus the system may be very close to spin disordered phase. Finally, attempts to fit the total energy data with Heisenberg-Kitaev model failed, suggesting that interactions beyond original Heisenberg-Kitaev model may be required to understand the magnetism of the system.
  
  
 
  To study the magnetic and electronic properties of Li$_2$RhO$_3$, we performed density functional based first principles calculations. In particular, we used the plane-wave basis and projected augmented wave method\cite{paw} as implemented in the Vienna Abinitio Simulation Package (VASP)\cite{vasp,vasp_paw}. The Perdew, Burke, and Ernzerhoff flavor (PBE)\cite{PBE} of generalized gradient approximation (GGA) were chosen to be the exchange-correlation functional. To ensure the convergence of total energy to 1 meV/cell, a high energy cut-off of 540 eV to the plane wave basis was chosen; while a dense $8\times5\times8$ $\Gamma$-centered K-grid was used to perform the Brillouin zone integration. The lattice constants and internal atomic positions were optimized so that the forces on individual atoms are smaller than 1 meV/\AA\ and internal stress less than 0.04 kbar.  The density of states (DOS) were calculated using a $16\times10\times16$ $\Gamma$-centered K-grid and tetrahedron method.


  Solution of the Heisenberg-Kitaev model suggested three possible AFM ground states for hexagonal lattices\cite{PhysRevLett.109.187201,PhysRevLett.110.097204,PhysRevB.87.041107}, which are also used to study the magnetic phases of Na$_2$IrO$_3$\cite{PhysRevLett.108.127203,PhysRevLett.108.127204} (FIG. \ref{fig:neel}-\ref{fig:stripy}). In our current study, we performed calculations for all three AFM states, as well as ferromagnetic (FM) state and non-magnetic (NM) state. The total energies of different magnetic phases are listed in TABLE \ref{tab:mag_energy}.

  Our calculations demonstrate that the SOC effect do play an essential role in determining the ground state magnetism. The FM phase is always the ground state if SOC is not included in the calculation, although the magnetic frustration is greatly enhanced with respect to increased $U$, as indicated by the reduction of the total energy difference between the FM phase and the second lowest phase (from $\sim$32 meV/Rh at $U=0$ eV to $\sim$2 meV/Rh at $U=4$ eV). With the SOC being correctly accounted for, the ground state of Li$_2$RhO$_3$ starts from the FM phase at $U=0$ eV to the zigzag AFM phase at $U=2$ eV, and eventually becomes stripy AFM phase at $U=4$ eV. It is worthy noting that once the AFM ground state is established, the total energy difference between the ground state and second lowest phase does not exceed 12 meV/cell (or 3 meV/Rh) (11 meV/cell at $U=2$ eV, 4 meV/cell at $U=3$ eV, and 7 meV/cell at $U=4$ eV). By comparison, the total energy difference between the NM phase and the highest magnetic phase is at least one order of magnitude larger (465 meV/cell at $U=2$ eV, 850 meV/cell at $U=3$ eV, 1253 meV/cell at $U=4$ eV). Such small energy difference between magnetic phases compared with the large magnetic energy manifests strong frustrations of magnetic interactions, and that the system may be very close to spin-glass or spin-liquid phases.

  
  

 We now turn to the electronic structure of Li$_2$RhO$_3$. Firstly, we compare the band structure of NM Li$_2$RhO$_3$ calculated without (Fig. \ref{fig:bs_nm}) and with (Fig. \ref{fig:bs_nm_soc}) SOC. Moderate band splitting due to SOC are normally expected for 4d-transition metals. However, for the current system, the major SOC effect is the band reordering around $\Gamma$ from $E_F$-0.5 eV to $E_F$+0.2 eV and from $E_F$-1.5 eV to $E_F$-0.8 eV. Nevertheless, a $\sim$ 86 meV band splitting due to SOC can still be identified around $\Gamma$ at $\sim E_F-1.0$ eV, which signals the strength of the SOC. Fig. \ref{fig:dos_nm} and \ref{fig:dos_nm_soc} show the total density of states (DOS) as well as the projected density of states (PDOS) of NM Li$_2$RhO$_3$ without and with SOC, respectively. Evidently, the electron states near Fermi level are contributed by the Rh-4d orbitals and O-2p orbitals, which hybridize significantly over a wide energy range. The $t_2g$ orbitals of rhodium due to the octahedral crystal field dominate the 4d contributions. At the NM phase, the system remains metallic even LDA+$U$+SOC is employed with $U$ up to 4 eV. 

 \begin{figure}[htp]
  \centering
  \subfigure[Stripy] {
    \rotatebox{270}{\scalebox{0.5}{\includegraphics{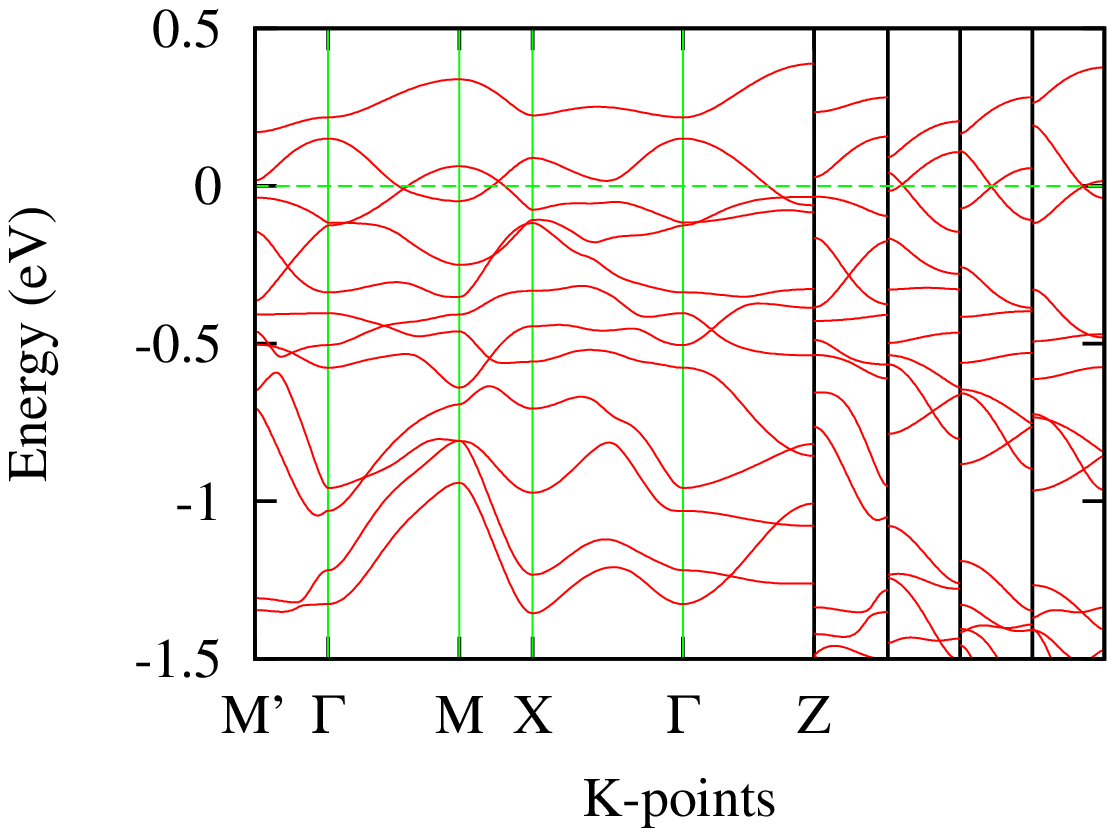}}}
    \label{fig:bs_col_nosoc}}
  \subfigure[Zigzag] {
    \rotatebox{270}{\scalebox{0.5}{\includegraphics{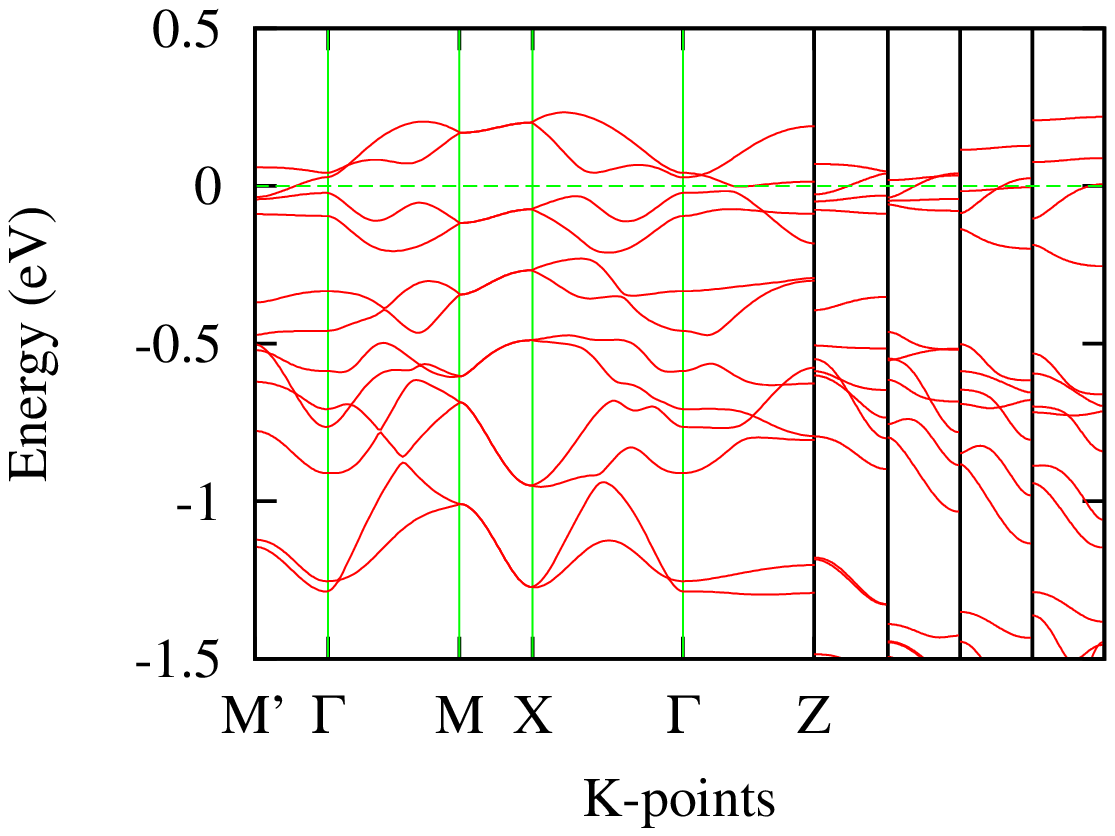}}}
    \label{fig:bs_bic_nosoc}}

  \caption{Band structure of (a) zigzag and (b) stripy phases without considering SOC effect at different $U$ level. The larger panel in each subfigure shows the band structure at $U=0$ eV; while the rest four smaller panels from left to right correspond to $U=$1, 2, 3, 4 eV, respectively. To save space, only the band structure from $M'$ to $\Gamma$ is shown for $U\neq0$ cases. \label{fig:bs_af_nosoc}}
 \end{figure}

 \begin{figure*}[htp]
  \centering
  \subfigure[Zigzag SOC+$U=0$ eV] {
    \rotatebox{270}{\scalebox{0.5}{\includegraphics{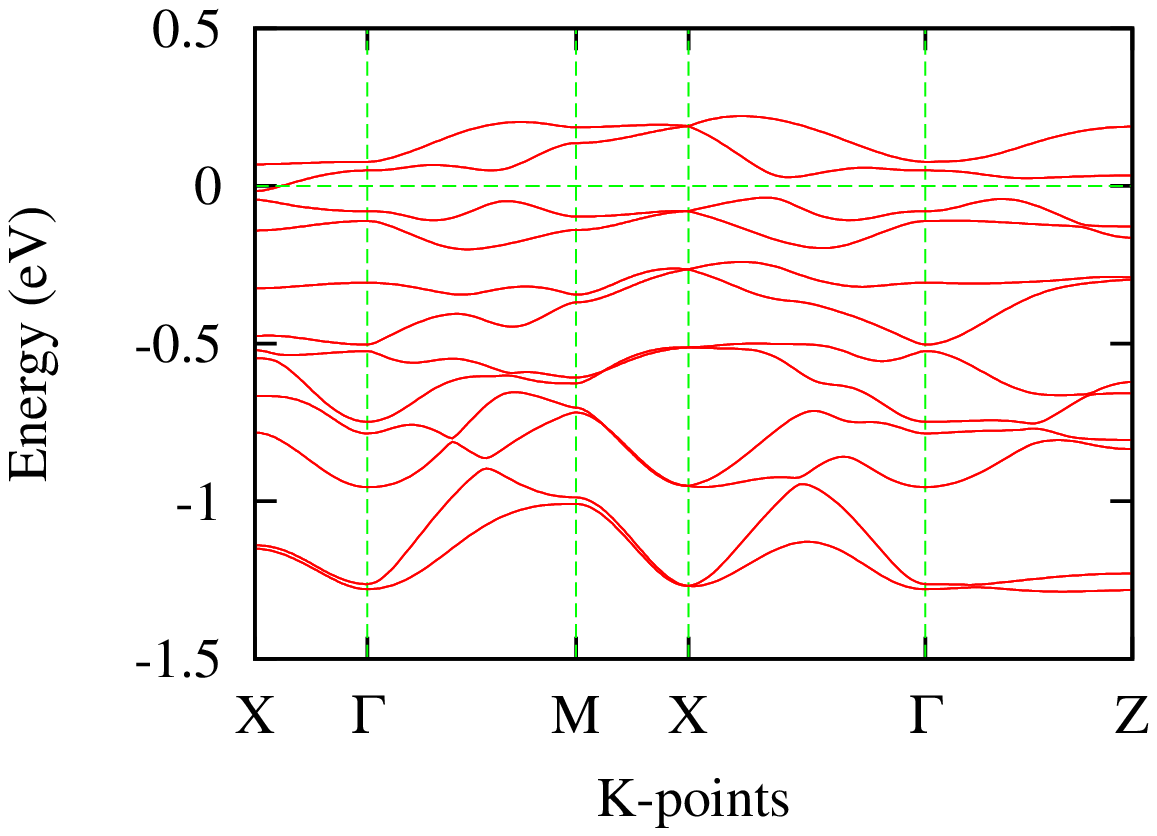}}}
    \label{fig:bs_socbic_u0}}
  \subfigure[Stripy SOC+$U=0$ eV] {
    \rotatebox{270}{\scalebox{0.5}{\includegraphics{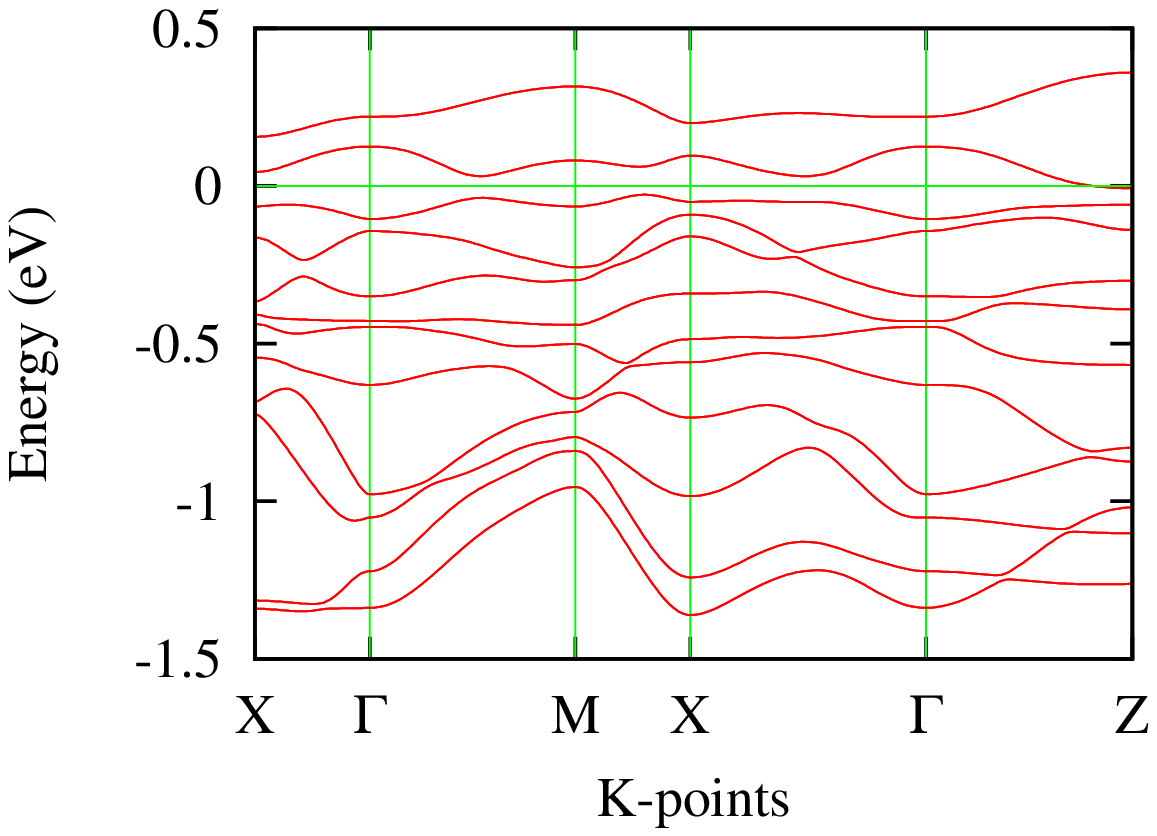}}}
    \label{fig:bs_soccol_u0}}
  \subfigure[Zigzag SOC+$U=2$ eV] {
    \rotatebox{270}{\scalebox{0.5}{\includegraphics{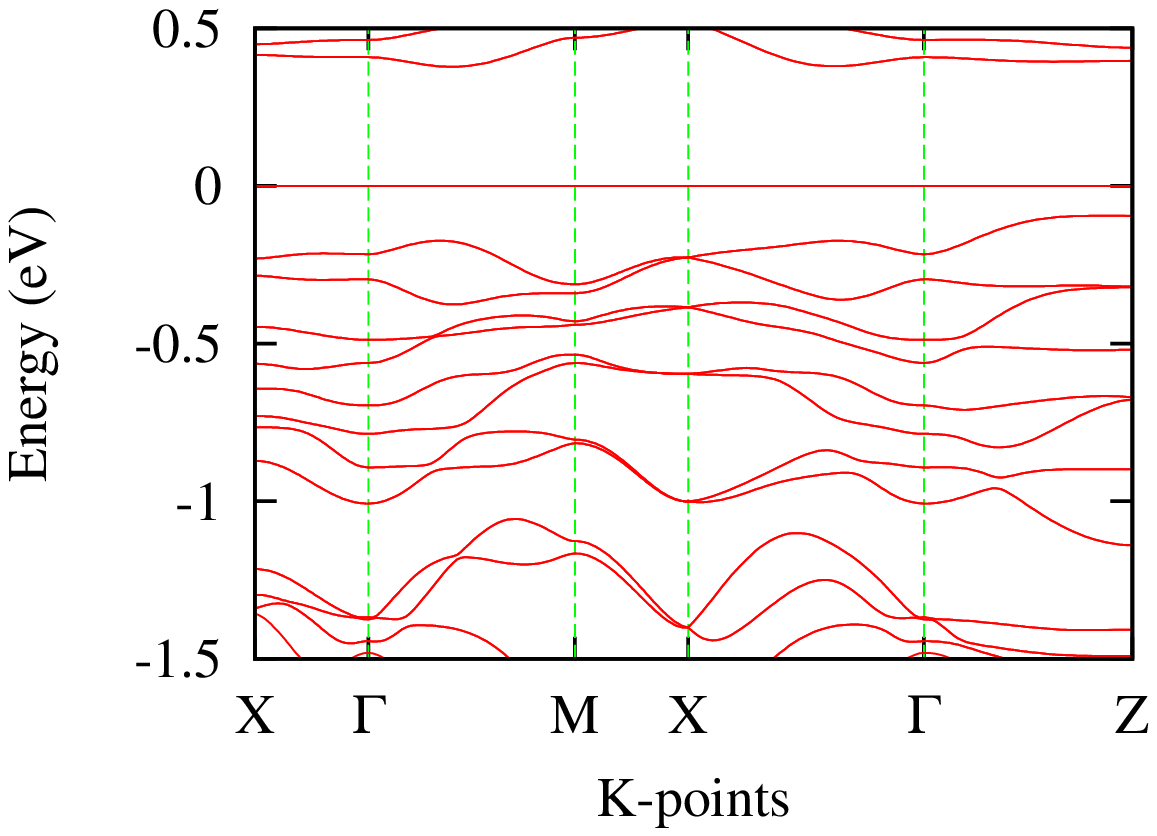}}}
    \label{fig:bs_socbic_u2}}
  \subfigure[Stripy SOC+$U=3$ eV] {
    \rotatebox{270}{\scalebox{0.5}{\includegraphics{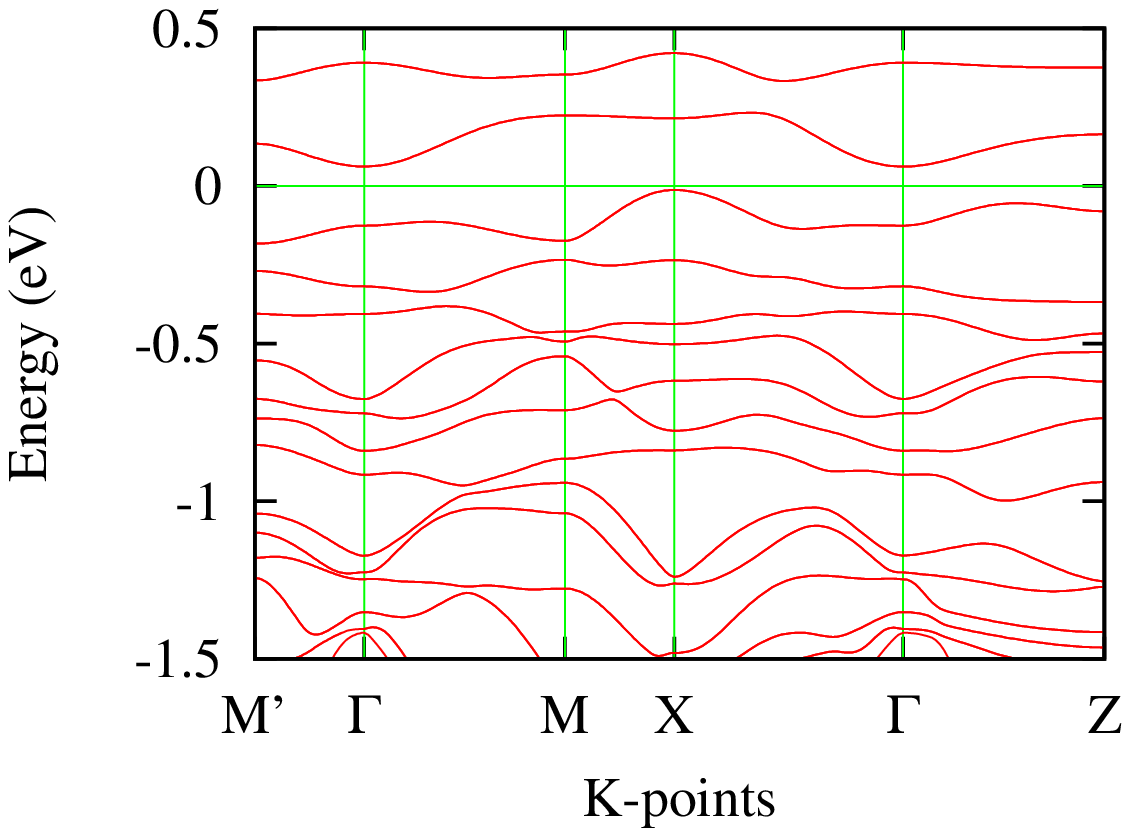}}}
    \label{fig:bs_soccol_u3}}

  \caption{Band structure of (a)(c) zigzag and (b)(d) stripy phases with SOC effect at different $U$ level. \label{fig:bs_af_soc}}
 \end{figure*}

 Secondly, we examine the electronic structure of Li$_2$RhO$_3$ at the zigzag- and stripy-AFM phases.\footnote{The N\'{e}el-AFM phase is not considered here because it is never the ground state in our calculation.} Although the AFM phases cannot be stabilized unless SOC is turned on, it is instructive to compare the electronic structure at different on-site $U$ without SOC first (FIG. \ref{fig:bs_af_nosoc}). The large on-site energy pushes away the bands near the Fermi level, causing significant reduction of electron states at $E_F$. However, the electron-electron correlation alone cannot drive the system insulating, since the band gap is not opened even with $U$ as high as 4 eV. With SOC, the system is also metallic at $U=0$ eV, although significant band reordering effect similar to the one observed in the NM case is also present in both AFM cases(FIG. \ref{fig:bs_socbic_u0} and \ref{fig:bs_soccol_u0}). However, a $\sim$10 meV energy gap immediately opens even with a small on-site energy $U=1$ eV in the zigzag-AFM phase. The gap grows to $\sim$75 meV and $\sim$130 meV, if the on-site energy $U$ is increased to 2 eV and 3 eV, respectively. Noticing that the experimentally observed $E_g$ is 78 meV, which is very close to the calculated value for zigzag-AFM phase at $U=2$ eV (FIG. \ref{fig:bs_socbic_u2}). The magnetic moment is found to be $\sim$ 0.4 $\mu_B$/Rh, twice the value for the one in the zigzag-AFM phase of Na$_2$IrO$_3$\cite{PhysRevLett.109.197201}. For the stripy-AFM phase, a relatively larger $U=2$ eV is required to open an diminishing energy gap of 3 meV. The gap expands to $\sim$62 meV and 117 meV for $U=3$ eV and $U=4$ eV, respectively. The magnetic moment is $\sim$0.2 $\mu_B$/Rh in this phase, similar to the value for the one in the stripy-AFM phase of Na$_2$IrO$_3$\cite{PhysRevLett.109.197201}. As the zigzag-AFM and stripy-AFM phases are energetically very close around $U=3$ eV ($\sim1$ meV/Rh), one cannot deduce the ground state magnetic ordering from the comparison between the calculation and experiment. Once again, the two phases severely competes with each other, and fluctuations (quantum or thermal) can easily drive the system into disordered spin states. Nevertheless, in each case the system is non-metallic with a small gap of 60$\sim$120 meV.
 
 We want to make some further remarks regarding the obtained results. The calculated band structure in all magnetic phases appear to be quite three-dimensional. In fact, the inter-layer Rh-Rh distances are $\sim$5.1 \AA\, still comparable with the intra-layer Rh-Rh distance of 3.03 \AA. Therefore, it is questionable whether two-dimensional Heisenberg-Kitaev model is applicable to such system. Indeed, when we attempted to fit the phase energy data to the classical results of the Heisenberg-Kitaev model\cite{PhysRevLett.109.187201}, the fitting failed for all $U$, as the fitted $J$ and $\alpha$ result in a wrong phase energy order. Therefore, it is possible that interactions beyond the original Heisenberg-Kitaev model is required to describe the system. Considering the three-dimensional feature of its electronic structure, we propose it is likely to be the inter-layer coupling $J_c$.
 
 In conclusion, we have performed first-principles study of Li$_2$RhO$_3$, comparing the electronic structure and energetics of different magnetic phases with or without SOC effect at $U$ ranging from 0 to 4 eV. Both SOC and on-site electron correlation are crucial in determining the electronic structure of the compound. Within the proper $U$ range, the compound is semiconducting with a small gap of $\sim$60 to $\sim$ 120 meV; and the energy difference between the competing magnetic phases are extremely small, suggesting highly frustrated magnetism in the system and that the system may be in close proximity to disordered spin state. Since the phase energies do not fit well with classical Heisenberg-Kitaev model results and its electronic structure is highly three-dimensional, we propose that inter-layer coupling $J_c$ may be required to explain the behavior of the system.

\begin{acknowledgments}
This work has been supported by the NSFC (No. 11274006, No. 11274084) and the NSF of Zhejiang Province (No. LR12A04003, Z6110033). All calculations were performed at the High Performance Computing Center of Hangzhou Normal University College of Science.
\end{acknowledgments}

\bibliography{Li2RhO3}
\end{document}